\documentclass[aps,prl,twocolumn,groupedaddress,superscriptaddress,showpacs]{revtex4}

\usepackage{graphicx}
\usepackage{dcolumn}
\usepackage{bm}
\usepackage[usenames]{color}

\usepackage[normalem]{ulem}

\begin{document}

\title{
Surface Scattering via Bulk Continuum States in the 3D Topological Insulator Bi$_{2}$Se$_{3}$
}

\author{Sunghun~Kim}
\affiliation{%
Graduate School of Science, Hiroshima University, 1-3-1 Kagamiyama, Higashi-Hiroshima 739-8526, Japan\\
}%

\author{M.~Ye}
\affiliation{%
Graduate School of Science, Hiroshima University, 1-3-1 Kagamiyama, Higashi-Hiroshima 739-8526, Japan\\
}%

\author{K.~Kuroda}
\affiliation{%
Graduate School of Science, Hiroshima University, 1-3-1 Kagamiyama, Higashi-Hiroshima 739-8526, Japan\\
}%

\author{Y.~Yamada}
\affiliation{%
Graduate School of Science, Hiroshima University, 1-3-1 Kagamiyama, Higashi-Hiroshima 739-8526, Japan\\
}%


\author{E.~E.~Krasovskii}
\affiliation{%
Departamento de F\'{\i}sica de Materiales UPV/EHU
and Centro de F\'{\i}sica de Materiales CFM - MPC
and Centro Mixto CSIC-UPV/EHU,
20080 San Sebasti\'an/Donostia, Basque Country, Spain
\\
}
\affiliation{%
Donostia International Physics Center (DIPC),
             20018 San Sebasti\'an/Donostia, Basque Country,
             Spain\\
}
\affiliation{%
IKERBASQUE, Basque Foundation for Science, 48011 Bilbao, Spain\\
}

\author{E.~V.~Chulkov}
\affiliation{%
Departamento de F\'{\i}sica de Materiales UPV/EHU
and Centro de F\'{\i}sica de Materiales CFM - MPC
and Centro Mixto CSIC-UPV/EHU,
20080 San Sebasti\'an/Donostia, Basque Country, Spain
\\
}
\affiliation{%
Donostia International Physics Center (DIPC),
             20018 San Sebasti\'an/Donostia, Basque Country,
             Spain\\
}

\author{K.~Miyamoto}
\affiliation{
Hiroshima Synchrotron Radiation Center, Hiroshima University, 2-313 Kagamiyama, Higashi-Hiroshima 739-0046, Japan\\
}%

\author{M.~Nakatake}
\affiliation{
Hiroshima Synchrotron Radiation Center, Hiroshima University, 2-313 Kagamiyama, Higashi-Hiroshima 739-0046, Japan\\
}%

\author{T.~Okuda}
\affiliation{
Hiroshima Synchrotron Radiation Center, Hiroshima University, 2-313 Kagamiyama, Higashi-Hiroshima 739-0046, Japan\\
}%

\author{Y.~Ueda}
\affiliation{
Kure National College of Technology, Agaminami 2-2-11, Kure 737-8506, Japan\\
}%

\author{K.~Shimada}
\affiliation{
Hiroshima Synchrotron Radiation Center, Hiroshima University, 2-313 Kagamiyama, Higashi-Hiroshima 739-0046, Japan\\
}%

\author{H.~Namatame}
\affiliation{
Hiroshima Synchrotron Radiation Center, Hiroshima University, 2-313 Kagamiyama, Higashi-Hiroshima 739-0046, Japan\\
}%

\author{M.~Taniguchi}
\affiliation{%
Graduate School of Science, Hiroshima University, 1-3-1 Kagamiyama, Higashi-Hiroshima 739-8526, Japan\\
}%
\affiliation{
Hiroshima Synchrotron Radiation Center, Hiroshima University, 2-313 Kagamiyama, Higashi-Hiroshima 739-0046, Japan\\
}%

\author{A.~Kimura}
\email{akiok@hiroshima-u.ac.jp}
\affiliation{%
Graduate School of Science, Hiroshima University, 1-3-1 Kagamiyama, Higashi-Hiroshima 739-8526, Japan\\
}%

\date{\today}

\begin{abstract}
We have performed scanning tunneling microscopy and differential tunneling
conductance ($dI/dV$) mapping for the surface of the three
dimensional topological insulator Bi$_{2}$Se$_{3}$.
The fast Fourier transformation applied to the $dI/dV$ image shows an
electron interference pattern near Dirac node despite the general belief 
that the backscattering is well suppressed in the bulk energy gap region.
The comparison of the present experimental result with theoretical surface and bulk band
structures shows that the electron interference occurs through the scattering between the
surface states near the Dirac node and the bulk continuum states.
\end{abstract}


\maketitle

%
\begin{figure}
\includegraphics[width=.8\columnwidth
]{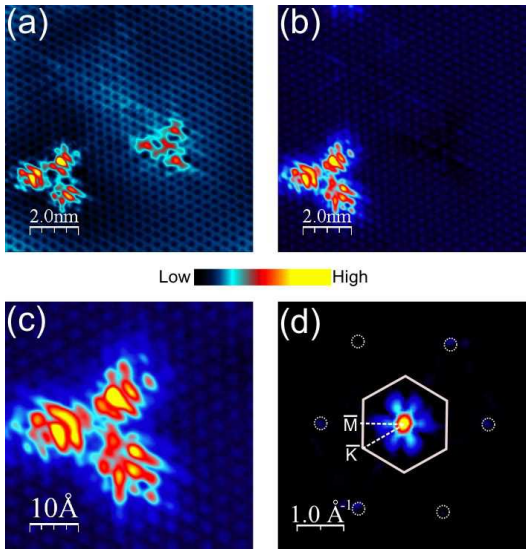} 
\caption{\label{fig:epsart1} (color online) (a) STM and
(b) differential conductance images of Bi$_{2}$Se$_{3}$ surface at a
sample bias voltage of $-300$~mV (10~nm$\times$10~nm). (c) Expanded
image of the clover-shaped pattern of (b) (5~nm$\times$5~nm). (d)
Fast Fourier transformation image of (c). All data sets are taken at
a temperature of 78K.}
\end{figure}
%

Recent discovery of topological insulators has aroused
great attention to the new state of quantum matter originating from
the surface state that forms a massless Dirac cone centered at time-reversal
invariant momentum~\cite{Kane_PRL_05,Bernevig_PRL_06,Bernevig_Science_06,Konig_Science_07,
  Hsieh_Nature_08, Hsieh_Science_08, Xia_NatPhys_09, Chen_Science_09, Kuroda_PRL_10, Fu_PRB_07,
  Zhang_NatPhys_09, Hsieh_Nature_09}.  Unlike the spin-degenerate
Dirac fermions in graphene, this novel electronic state possesses
helical spin textures protected by time-reversal symmetry, which creates
the potential for spintronics devices without heat dissipation and for
quantum computation applications. The topological insulator phase has been predicted
to exist in a number of three-dimensional materials, Bi$_{1-x}$Sb$_{x}$~\cite{Fu_PRB_07}, 
Bi$_{2}$Se$_{3}$, Bi$_{2}$Te$_{3}$, and Sb$_{2}$Te$_{3}$~\cite{Zhang_NatPhys_09}. 
The spin-helical surface states have been experimentally observed 
by angle-resolved photoelectron spectroscopy (ARPES)~\cite{Hsieh_Nature_08, Xia_NatPhys_09, Chen_Science_09, Kuroda_PRL_10}, 
also with spin resolution~\cite{Hsieh_Science_08, Hsieh_Nature_09, Nishide_PRB_10}.

Among the binary chalcogenides, Bi$_{2}$Se$_{3}$ is regarded as the most
promising candidate owing to the single ideal Dirac cone residing in a wide band
gap (0.3~eV). Therefore, significant effort has been applied towards the
realization of spintronics devices using the quantum topological transport
in Bi$_{2}$Se$_{3}$~\cite{Checkelsky_PRL_09,Analytis_PRB_10, Eto_PRB_10, Butch_PRB_10}.
The magnetotransport measurements showed Shubnikov-de
Haas oscillations originating from the 3D Fermi surface, but no surface contribution to
conduction has been observed in 3D topological insulators even at low carrier
density~\cite{Analytis_PRB_10, Eto_PRB_10, Butch_PRB_10}.

These observations pose the question of what limits the surface electron
conduction in Bi$_{2}$Se$_{3}$.
Band structure calculations of Ref.~\cite{Xia_NatPhys_09} predict
the Dirac point of the surface state in Bi$_{2}$Se$_{3}$ to be located below or close
to the bulk valence band maximum, which may open a scattering channel from the surface
states to bulk continuum states~\cite{Xia_NatPhys_09}. In this Letter, we
present scanning tunneling microscopy evidence of the near-surface scattering of the spin polarized surface states 
at the Dirac point in Bi$_{2}$Se$_{3}$ into spin-degenerate bulk continuum states.

The scanning tunneling microscopy (STM) and spectroscopy (STS) experiments were conducted
at 78K and at 4.5K in an ultrahigh vacuum with a base pressure of 1$\times$10$^{-10}$~mbar
using a low-temperature scanning tunneling microscope (LT-STM, Omicron NanoTechnology).
The STM images were acquired in the constant-current mode with a bias voltage $V_{s}$ applied
to the sample. The 
$dI/dV$ images were obtained simultaneously
by recording the STM images using a lock-in technique. Bi$_{2}$Se$_{3}$ samples were cleaved
{\it in situ} at room temperature in an ultrahigh vacuum to minimize the surface contamination.
For comparison, a photoemission spectrum was measured
with a hemispherical photoelectron analyzer (VG-SCIENTA R4000) at Hiroshima Synchrotron Radiation Center, Hiroshima University.
The crystal structure of Bi$_{2}$Se$_{3}$ consists of weakly 
bonded 5-layer (quintuple layer) blocks composed of Se-Bi-Se-Bi-Se atomic layers.
Therefore, the topmost Se layer is exposed to the ultra-high vacuum.

%
\begin{figure}
\includegraphics[width=.8\columnwidth
]{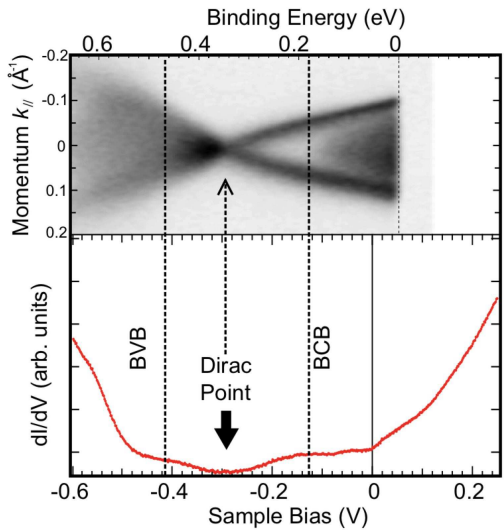} 
\caption{\label{fig:epsart2} (color online) Experimental
local density of states of Bi$_{2}$Se$_{3}$ surface acquired at
4.5~K ({\it lower}). Corresponding energy dispersion curve measured 
by 
ARPES for the 
surface state of Bi$_{2}$Se$_{3}$ is shown for comparison ({\it upper}).
Both are shown in the same energy scale and their energy axes are fixed at the Dirac point.}
\end{figure}
%

The STM topographic image and the 
$dI/dV$ map acquired at 78K for the sample bias voltage $V_{s}=-300$~mV are shown in Figs.~1(a) 
and 1(b). We find two different defect images with clover- and triangular-shaped
patterns superimposed on the atomically resolved periodic structure spreading over the entire surface
area in the topographic STM image. In the $dI/dV$ image of Fig.~1(b) the clover-shaped pattern can still
be seen, but not the triangular one.
If we take a close look at the clover-shaped pattern
in Fig.~1(c), we find the intensity modulation near the defect
with a much longer wavelength than the surface lattice constant. 
A similar pattern has been observed in the pioneering STM study~\cite{Urazhdin_PRB_02}
and identified as a defect state induced by an excess Bi atom
substituting the bottom Se site of the topmost quintuple layer. 
It was suggested that a resonance state is formed inside the bulk
valence band due to the defect. 
Since the concept of topological insulator was not established at that time,
we now need to revisit and reconsider the origin of the
peculiar pattern in the STM image by taking into account the presence of the
Dirac fermions with helical spin texture.

In Fig.~1(d) we show the fast Fourier transformation (FFT) of the $dI/dV$ image. 
Here the surface Brillouin zone (SBZ) boundaries are
drawn from the obtained sharp FFT spots originated from the surface periodicity.
The FFT image unambiguously tells us that an electron
interference takes place as is seen from the flower shaped intensity distribution (Fig.~1(d)).
The maximal intensity is seen to be along the
$\bar{\Gamma}\bar{M}$ direction of the scattering vector $\mathbf{q}$, 
exhibiting a node in the $\bar{\Gamma}\bar{K}$ direction.

In order to further reveal the electron interference effect around the defects in
Bi$_{2}$Se$_{3}$, we have performed the experiment at a liquid He temperature (4.5~K).
Figure~2 shows the experimental local density of states (LDOS) 
measured as the differential tunneling conductance $dI/dV$ (STS) spectrum of Bi$_{2}$Se$_{3}$ surface.
 We find a V-shaped STS spectrum with a minimum at $-300$~mV 
relative to the Fermi energy ($E_{F}$), which corresponds to the Dirac
point (DP).  
The slope of STS spectrum is seen to become much steeper below
$-500$~mV and above $E_{F}$.
By comparing the present STS spectrum with the reported ARPES energy dispersion~\cite{Kuroda_PRL_10}, 
we find the bulk valence band (BVB) onset at $-500$~meV
leading to the enhancement of LDOS below this energy. The shoulder 
at $-100$~meV coincides with the onset of bulk conduction band (BCB) as shown in Fig.~2.

%
\begin{figure}
\includegraphics[width=\columnwidth
]{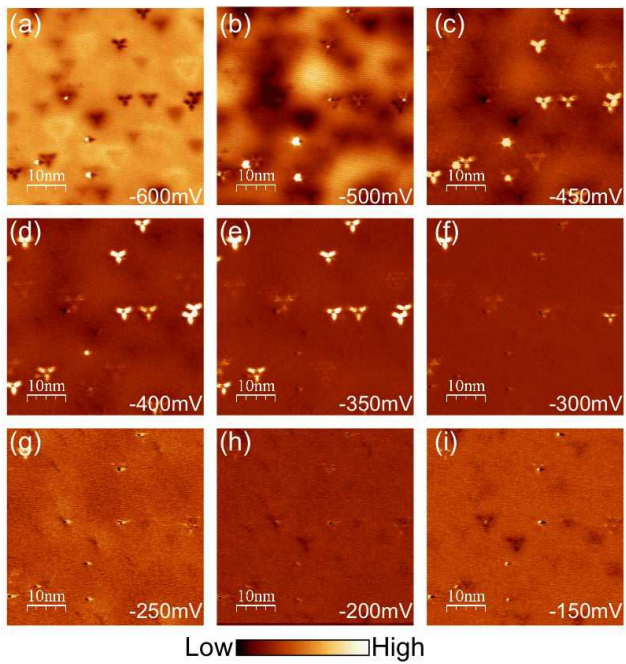} 
\caption{\label{fig:epsart3} (color online) (a)-(i)
Differential conductance images of Bi$_{2}$Se$_{3}$ surface at 4.5K
for several sample bias voltages from $-600$~mV to $-150$~mV in the
area of 50~nm$\times 50$~nm.}
\end{figure}
%

Figures 3(a)-3(i) represent the differential tunneling conductance
$dI/dV$ images from $V_{s}=-600$ to $-150$~mV in the 50~nm$\times 50$~nm
scanning area. Here the clover-shaped defect again appears with a brighter contrast with respect
to the rest part of the images in a limited bias window between $-450$ and $-300$~mV (Figs.~3(c)-3(f)), 
which is close to and a little below the DP. 
However, it loses contrast at higher voltages ($V_{s} > -250$~mV) and
becomes even darker at $V_{s} < -500$~mV (Figs.~3(a) and 3(b)).
%

\begin{figure*}
\includegraphics
[width=.9\textwidth] {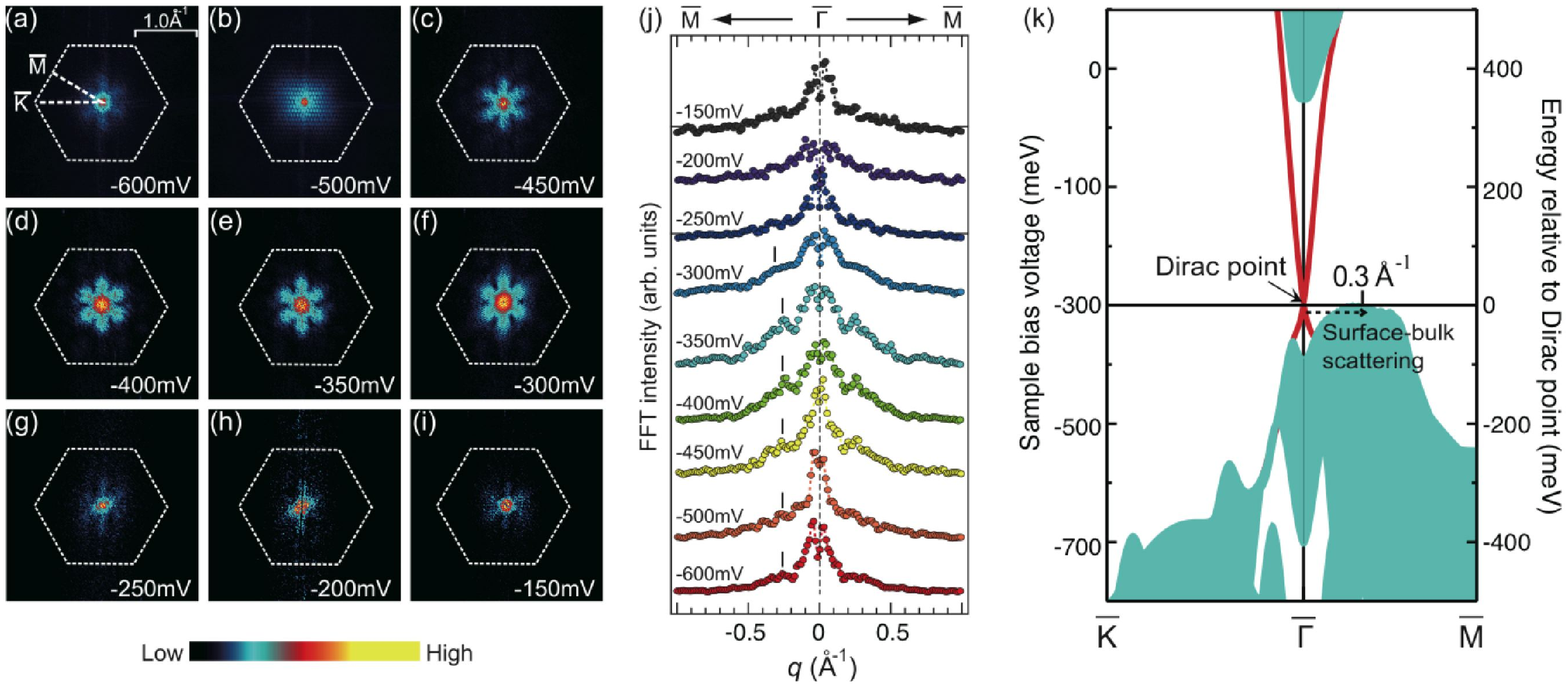}
\caption{\label{fig:wide} (color online) (a)-(i) Fast Fourier transformation 
of the $dI/dV$ images in Figs.~3(a)-3(i) for respective bias
voltages. (j) Cross-sectional profiles of the FFT images for (a)-(i)
along $\bar{\Gamma}\bar{ M}$. (k) Theoretical surface state
dispersion (solid lines) and $\mathbf k_\parallel$ projected
bulk band structure (shaded area).
An arrow shows possible surface to bulk scattering channels.}
\end{figure*}
%

To quantify the observed electron interference effect and determine
the relevant $\mathbf{q}$ values, the FFT has been applied to the $dI/dV$ 
images in Figs.~4(a)-4(i). At first, we observe a characteristic continuous 
intensity distribution with sixfold symmetry inside the SBZ for $V_{s}=-450$,
$-400$, $-350$, and $-300$~mV, for which the clover-shaped patterns are observed in
the corresponding $dI/dV$ images (Figs.~4(c)-4(f)).
Of the four FFT images, the intensity is the most prominent at $V_{s}=-400$ and $-350$~mV, 
which coincides with the most distinct contrast in the clover-shaped pattern in Fig.~3(d) and 3(e).
The characteristic six-fold pattern disappears above $-250$~mV, where only a weak
intensity remains around the SBZ center (Figs.~4(g)-4(i)). Below $-450$~mV, the FFT
patterns become less anisotropic and their intensity is weakened as the bias
voltage decreases.
It follows from the FFT analysis that the anisotropic interference 
is strongest between $-450$ and $-300$~mV, i.e., close to and slightly 
below the Dirac point.

To gain deeper insight into the origin of the unexpected 
interference pattern,
the cross-sectional FFT patterns traced along $\bar{\Gamma}\bar{ M}$ direction
are shown in Fig.~4(j). For the whole range of sample bias voltages, we observe
the largest intensity at $q$=0, which might originate from several
extrinsic factors, such as a background noise whose modulation length would be much
longer than the characteristic wavelength of the surface lattice periodicity and the
electron interference. 

On the topological insulator surface, the spins are locked with momenta, giving 
rise to helical Dirac fermions with lifted spin degeneracy. Owing to the strictly 
antiparallel spins at $-\mathbf{k}$ and $\mathbf{k}$, 
surface electrons are protected from backscattering, and no standing
wave pattern arising from the interference with the backscattered
wave can be observed for a single Dirac cone.
On the other hand, there might be other scattering channels 
within the surface state cone with a finite parallel spin component, which could 
contribute to the scattering with small $\mathbf{q}$ values of less than 0.2~\AA$^{-1}$
(the maximal radius of the cone is 0.1~\AA$^{-1}$). 
Although a detailed structure of surface state scattering is not resolved in the present
data, the dominant contribution around $\mathbf{q}$=0 apparently originates from the 
extrinsic effect because a finite intensity at $\mathbf{q}$=0 persists and even enhances 
below $-500$~mV, where the surface states are merged into the bulk valence band.

Let us now consider broad humps centered at $q=\pm 0.3$~\AA$^{-1}$,
which are 
especially pronounced in the sample bias range from
$-450$ to $-300$~mV
as indicated by vertical bars in Fig.~4(j). The intensities in this cross sectional
FFT profile are spread over a wide interval $-0.5<q<0.5$~\AA$^{-1}$.
Note that the FFT pattern is anisotropic, i.e., the intensity
vanishes in the $\bar{\Gamma}\bar{ K}$ direction, see Figs.~4(c)-4(f).
The observed magnitude of the  scattering vector, $|q|\sim0.3$~\AA$^{-1}$,
is too large to be explained by a scattering within the Dirac cone, especially near
the DP. 
Besides, the profile is too broad to be explained by a scattering between the surface states
well localized in $\mathbf k_\parallel$.
Also note that the intensity maxima of the hump are
nearly independent of the bias voltage.
Another important observation is that the hump almost disappears
at 50~mV above the DP, which also excludes its origin from
the surface state scattering.  Since the scattering occurs at the near-surface
defect, we need to consider both surface and bulk band structure to figure out a
possible scattering channel other than the surface states of the Dirac cone.

We have performed a first principles calculation of the surface states with a
slab of seven formula units with a two-component self-consistent (within the
local density approximation) full-potential augmented plane wave method~\cite{Krasovskii_PRB_99}.
According to the theoretical band structure
the DP is energetically located slightly below the bulk valence
band maximum in the middle of the $\bar{\Gamma}\bar{ M}$ interval,
whereas around the $\bar{\Gamma}$ point the bulk band gap widens, so the states
of the Dirac cone are energetically well separated from the bulk states with
small $\mathbf k_\parallel$ vectors (Fig.~4(k)).
The interference between a surface state with the lateral Bloch vector
$\mathbf k_\parallel^{s}$ and a bulk state with $\mathbf k_\parallel^{b}$
caused by the scattering at the point defect results in spatial oscillations
of LDOS with the wave vector $\mathbf q=\mathbf k_\parallel^{s}-\mathbf k_\parallel^{b}$.
The possible $s\to b$ transitions at the DP energy are seen to be centered
at $q\approx 0.3$~\AA$^{-1}$ in the $\bar{\Gamma}\bar{ M}$ direction. At
the same time, in the $\bar{\Gamma}\bar{ K}$ direction such transitions
are forbidden in a wide energy interval: the valence band maximum is located
much lower in energy than in the $\bar{\Gamma}\bar{ M}$ direction. This
striking anisotropy explains the negligible FFT intensity along
$\bar{\Gamma}\bar{ K}$ line, which results in the flower-shaped distribution in Fig.~4.
Since the bulk states are continuously distributed over the
surface perpendicular momentum the interference pattern broadens, leading
to a smeared hump in Fig.~4(j).
Between the valence band maximum and the conduction band
minimum there exists a large portion of the upper cone that 
does not energetically overlap with bulk states. This is consistent with
the abrupt disappearance of the hump above $V_{s}=-200$~mV (Fig.~4(j)).
This means that the surface-bulk scattering is strongly 
suppressed above the DP, which supports our interpretation of the
hump at $q\approx 0.3$~\AA$^{-1}$ as due to the scattering into
bulk states.
%
%
A similar scattering may also occur via the bulk conduction band for $V_{s} >$ -100~meV with a smaller scattering vector $|q|\le0.2$~\AA$^{-1}$, which is beyond the resolution limit of this measurement. Most probably, it is this process that gives rise to the triangular-shaped pattern in Fig.1, as it appears in the topographic image and disappears in the dI/dV map.
%

In summary, we have shown by scanning tunneling microscopy and
differential tunneling conductance measurements that a strong surface
state -- bulk continuum scattering occurs near the Dirac node at the
surface of the three dimensional topological insulator
Bi$_{2}$Se$_{3}$, contrary to the belief that the surface state in the
bulk energy gap around $\bar{\Gamma}$ point
is protected from scattering~\cite{Xia_NatPhys_09}.  
This may have implications for the construction of spintronic
devices. In particular, it has been proposed to tune the DP at the top 
and the bottom surfaces of a thin film below and above the chemical potential 
in the dual-gate device configuration~\cite{Yazyev_PRL_10}. To suppress
the scattering one would have to separate both the lower and the upper 
Dirac cone from the bulk states. Also, in order to realize the anomalous 
Hall effect by an appropriate magnetic doping, one must precisely tune the 
chemical potential right at the Dirac point to achieve the surface insulating 
state by opening the energy gap~\cite{Ru_Science_10}.
Thus, we expect the present results to stimulate future band
engineering in order to isolate the Dirac surface state from the bulk
continuum and facilitate topological quantum transport.

We thank Shuichi Murakami of Tokyo Institute of Technology for
valuable discussion.  The ARPES measurement was performed with the
approval of the Proposal Assessing Committee of HSRC (Proposal
No. 09-A-52).  This work was financially supported by the JSPS
Grant-in-Aid for Scientific Research (B) 20340092.

\end{document}